\documentclass[11pt,a4paper]{article}
\usepackage{jcappub}
\pdfoutput=1

\ifx\pdfoutput\undefined
\usepackage[dvips,bookmarks=false]{hyperref}	% This is for arXiv.org
\else
\usepackage{hyperref}	% This is for pdftex
\fi
\hypersetup{colorlinks,bookmarksopen,bookmarksnumbered,citecolor=blue,
linkcolor=black,pdfstartview=FitH,urlcolor=blue}

\usepackage{graphicx}
\usepackage{amsmath}
\usepackage{amssymb}

\usepackage{color}

\notoc

\title{Comment on ``Strong Evidence for the Normal Neutrino Hierarchy''}

\def\okc{Oskar Klein Centre for Cosmoparticle Physics,
  Department of Physics, Stockholm University, SE-10691 Stockholm, Sweden}
\def\kit{Institut f\"ur Kernphysik, Karlsruhe Institute of Technology, 76021 Karlsruhe, Germany}
\def\mellon{McWilliams Center for Cosmology, Department of Physics,
  Carnegie Mellon University, Pittsburgh, PA 15213, USA}
\def\lbnl{Lawrence Berkeley National Laboratory (LBNL),
  Physics Division, Berkeley, CA 94720-8153, USA}
\def\berkley{Berkeley Center for Cosmological Physics, University of California, Berkeley, CA 94720, USA}
\def\ific{Instituto de Fisica Corpuscular (IFIC), Universidad de Valencia-CSIC, E-46980, Valencia, Spain}
\def\michi{Michigan Center for Theoretical Physics, Department of Physics,
  University of Michigan, Ann Arbor, MI 48109, USA}
\def\infn{Instituto Nazionale di Fisica Nucleare (INFN), Sezione di Ferrara, I-44122 Ferrara, Italy}
\def\aarhus{Department of Physics and Astronomy, Aarhus University, Ny Munkegade, DK-8000 Aarhus C, Denmark}

\author[a]{T.~Schwetz,}
\author[b,c]{K.~Freese,}
\author[b]{M.~Gerbino,}
\author[d]{E.~Giusarma,}
\author[e]{S.~Hannestad,}
\author[f]{M.~Lattanzi,}
\author[g]{O.~Mena,}
\author[b]{S.~Vagnozzi}

\affiliation[a]{\kit}
\affiliation[b]{\okc}
\affiliation[c]{\michi}
\affiliation[d]{\mellon; \lbnl; \berkley}
\affiliation[e]{\aarhus}
\affiliation[f]{\infn}
\affiliation[g]{\ific}

\emailAdd{schwetz@kit.edu, martina.gerbino@fysik.su.se}

\abstract{In the preprint arxiv:1703.03425 ``strong evidence'' for the
  normal neutrino mass ordering is claimed. The authors obtain
  Bayesian odds of 42:1 in favour of the normal ordering. Their
  conclusion is based on adopting a flat logarithmic prior for the
  three neutrino masses. Such an assumption favours a hierarchical
  spectrum for the masses, which is much easier to accommodate for the
  normal mass ordering, and hence their prior assumption makes the
  inverted ordering much less likely {\it a priori}. We argue that the
  claimed ``evidence'' for normal ordering is almost entirely driven by the
  adopted prior and not due to the data itself.
}

\begin{document}
\maketitle

\newpage

In a recent pre-print \cite{Simpson:2017qvj} entitled ``Strong
Evidence for the Normal Neutrino Hierarchy'' by F.~Simpson,
R.~Jimenez, C.~Pena-Garay, and L.~Verde (SJPV in the following) a
Bayesian analysis of data from neutrino oscillation experiments and
cosmology has been performed. SJPV obtain Bayesian odds of 42:1 in
favour of the normal hierarchy (or ordering), a much stronger result
than obtained by similar recent Bayesian analyses in
Refs.~\cite{Hannestad:2016fog,Gerbino:2016ehw,Vagnozzi:2017ovm} as
well as the $\chi^2$-based analysis from Ref.~\cite{Capozzi:2017ipn}.

In this note we want to clarify that the result of SJPV is a
consequence of the assumed Bayesian prior, {\it i.e.,} the subjective
belief of the authors before data is used. SJPV assume that the three
neutrino masses have a flat probability distribution in logarithmic scale. This
assumption favours a hierarchical spectrum and makes the inverted
ordering (where the two heavier masses are quasi-degenerate) much less
likely. On top of this assumption, SJPV impose a so-called
hyper-prior, which limits the available range for the three masses in
logarithmic scale. The effect of this hyper-prior is that also a
quasi-degenerate spectrum becomes likely if the hyper-prior is chosen
to have a narrow width.
Hence, the particular choice of their prior favours either a
hierarchical or a quasi-degenerate spectrum. For normal ordering, both
options are available, whereas in case of inverted ordering, only a
quasi-degenerate spectrum is likely under these prior plus hyper-prior
assumptions (this is visible in Fig.~3 of
\cite{Simpson:2017qvj}). Once the cosmological bound on the sum of
neutrino masses is imposed, a quasi-degenerate spectrum is
disfavoured, which leaves no room for the inverted ordering and leads
to the strong preference for the normal hierarchical spectrum.

\bigskip

In summary, the claimed preference for normal ordering is strongly
driven by the assumed prior. The work of SJPV quantifies the following
statement: if a hierarchical spectrum is assumed (flat prior in the
logarithm of the three masses), then a normal ordering is preferred
over inverted. The assumption of a logarithmic distribution of the
masses may have some appeal (for instance having in mind the masses of
other fermions of the Standard Model). However, the conclusion reached
by SJPV should not be confused with the question of whether {\it data}
by themselves can distinguish between normal and inverted ordering
when assuming that a priori they are equally likely.

It is this latter question which has been addressed in
Refs.~\cite{Hannestad:2016fog,Gerbino:2016ehw,Vagnozzi:2017ovm}, where
a Bayesian prior has been chosen in such a way, that the odds for
normal versus inverted ordering are $\approx 1:1$ before cosmological
data is used. Once data from cosmology is added we will learn to what
extent the {\it data} can disfavour inverted ordering, independent of
any {\it a priori} assumption of how one expects masses to be
distributed. In this way, in
Refs.~\cite{Hannestad:2016fog,Gerbino:2016ehw,Vagnozzi:2017ovm} odds
in favour of normal ordering of between approximately 2:1 and 3:1
(depending on the used data) are obtained, in qualitative agreement
with Ref.~\cite{Capozzi:2017ipn}.\footnote{Note that when a similar
  data set leading to odds of 3:1 in \cite{Vagnozzi:2017ovm} would be
  used in an analysis as in SJPV, odds of about 90:1 or higher would
  be obtained.}

Let us stress that the sensitivity of current cosmological data is
entirely based on the available parameter spaces for the two mass
orderings, {\it i.e.}, volume effects in the Bayesian
language. Therefore it is important to specify priors such that the
analysis addresses the question one is interested in. The prior chosen
by SJPV mixes the question of the mass ordering with an assumption
about the distribution of the individual masses, whereas the analyses
of Refs.~\cite{Hannestad:2016fog,Gerbino:2016ehw,Vagnozzi:2017ovm}
investigate purely the question of normal versus inverted ordering,
irrespective of the distribution of individual masses.

%\bibliographystyle{JHEP_improved.bst}
%\bibliography{./refs}

\providecommand{\href}[2]{#2}\begingroup\raggedright\endgroup

\end{document}